\documentclass[nologo,11pt,a4paper]{ETHpaper}
\usepackage{graphicx, epsfig, amsmath, amssymb,color,wasysym}
\usepackage[square,numbers,sort&compress]{natbib}
\usepackage{caption}
\usepackage{subcaption}
\usepackage{longtable}
\usepackage{booktabs}
\usepackage{mathtools}
\usepackage{cancel}
\usepackage{ragged2e}

\hypersetup{
    colorlinks=false,
    pdfborder={0 0 0},
 }

\usepackage{soul,color} 
\usepackage{pdfpages}

\begin{document}

\title{The digital traces of bubbles: feedback cycles between socio-economic signals in the Bitcoin economy}
\titlealternative{Feedback cycles between socio-economic signals in the Bitcoin economy}

\author{ David Garcia\footnote{Corresponding author:
    \url{dgarcia@ethz.ch}}, Claudio J. Tessone, Pavlin Mavrodiev, and
  Nicolas Perony}
\authoralternative{D. Garcia, C.J. Tessone, P. Mavrodiev, N. Perony}

\address{Chair of Systems Design, ETH Zurich, Weinbergstrasse 56/58, 8092 Zurich, Switzerland}
\www{\url{http://www.sg.ethz.ch}}
\makeframing
\maketitle

\begin{abstract} 
What is the role of social interactions in the creation of price bubbles?
Answering this question requires obtaining collective behavioural traces generated by the activity of a large number of actors.
Digital currencies offer a unique possibility to measure socio-economic signals from such digital traces.
Here, we focus on Bitcoin, the most popular cryptocurrency.
Bitcoin has experienced periods of rapid increase in exchange rates (price) followed by sharp decline; we hypothesise that these fluctuations are largely driven by the interplay between different social phenomena.
We thus quantify four socio-economic signals about Bitcoin from large data sets: price on on-line exchanges, volume of word-of-mouth communication in on-line social media, volume of information search, and user base growth.
By using vector autoregression, we identify two positive feedback loops that lead to price bubbles in the absence of exogenous stimuli: one driven by word of mouth, and the other by new Bitcoin adopters.
We also observe that spikes in information search, presumably linked to external events, precede drastic price declines.
Understanding the interplay between the socio-economic signals we measured can lead to applications beyond cryptocurrencies to other phenomena which leave digital footprints, such as on-line social network usage.
\end{abstract}

\centerline{\small {\bf Key index words}: Social interactions, Bubbles, Socio-economic signals, Bitcoin}

\section{Introduction}
Bitcoin \cite{nakamoto2008}, the best-known cryptographic currency, draws as many harbingers of imminent failure \cite{hadas2013,blackstone2014} as heralds of long-term success as a mainstream currency \cite{fontevecchia2013}. Throughout its 5-year existence, it has been the subject of growing attention, due in part to its rapidly increasing and very volatile exchange rate to other currencies. Amidst the hype surrounding the cryptocurrency, it is difficult to recognise which factors participate in its growth, and influence its value. Bitcoin's decentralised structure, based on the contribution of its users rather than a central authority, implies that the dynamics of its economy may be strongly driven by social factors, which are composed of interactions between the actors of the market. This paper reveals the interdependence between social signals and price in the Bitcoin economy, namely a social feedback cycle based on word-of-mouth effect and a user-driven adoption cycle.

The Bitcoin economy is indeed growing at a staggering speed: the market value of all bitcoins in circulation went from about USD \$277K when bitcoins were first publicly traded in July of 2010, to over USD \$14B in December of 2013, hence a fifty-thousand-fold increase in that period \cite{blockchaininfo2013}. This came along with a stark increase in public interest, as shown by Internet search data: during the same period, the volume of Bitcoin-related searches on the Google search engine grew by over 10'000\% \cite{googletrends2013}.
Our hypothesis is that this growth is driven by the online actions and interactions of individual users, which leave traces of their activity. How can we use these digital traces to capture the link between market dynamics and public interest? In this paper, we provide evidence that this feedback can be understood by incorporating two types of signals: price and social information.
The influence of the first was investigated through the foundational work of Fama \cite{Fama1969} and later Grossman \cite{grossman1976}, who demonstrated that economic agents rapidly integrate common sources of information to assign a price to a good, including price itself.
The role of purely social information for price formation was first studied by Bikhchandani \cite{bikhchandani1992}, who showed that imitation is a rational strategy in periods of large volatility or in the absence of other sources of information.

\begin{figure}[ht!]
\centering
\includegraphics[width=\textwidth]{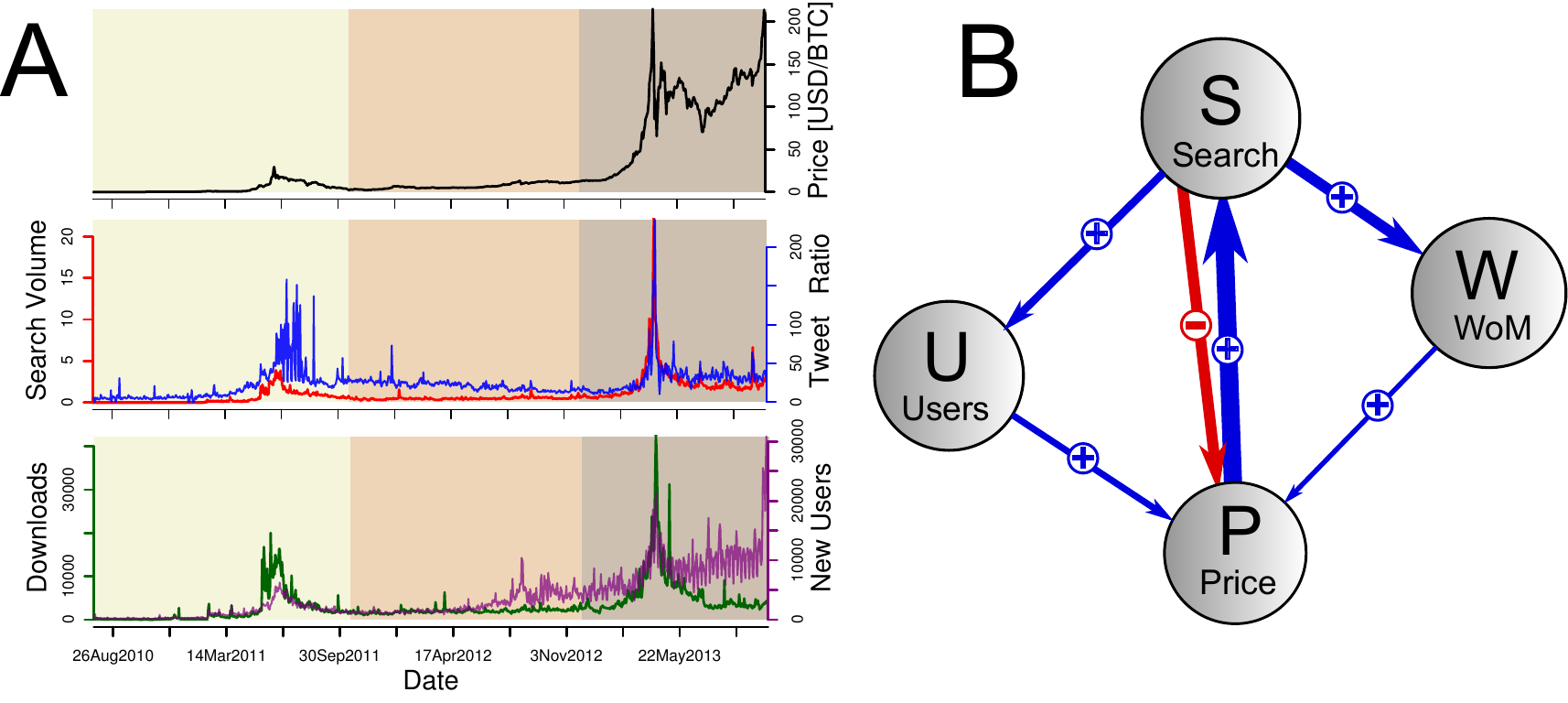}

\caption{ {\bf A.} Time series of price (black) on Mt. Gox (top), number of Bitcoin-related tweets per million tweets (blue), search volume (red) on Google (middle), and number of new users in the Bitcoin network (purple) and number of downloads of the bitcoin client (green) in SourceForge (bottom). Differently coloured backgrounds denote the three periods of our analysis. {\bf B.} Feedback diagram for the variables of our analysis. Increasing Bitcoin prices create collective attention through search volumes, which in turn triggers word of mouth about Bitcoin, leading to higher prices. A similar loop exists with the amount of users in the Bitcoin economy. Very high search volumes serve as an indicator of information search patterns before users sell their bitcoins, lowering the price.  Arrows connecting variables have widths proportional to the vector autoregression results of Table~\ref{tab:VAR}. }
\label{fig:timeSeries}
\end{figure}

In the Bitcoin economy, the fixed supply and predictable scarcity, both independent of the user base, create a strong link between public interest, user adoption, and price (illustrated in the time series of Fig.~\ref{fig:timeSeries}a). Following from our hypothesis on the role of social interactions, a key issue is characterising the effect of social influence \cite{Saavedra2011,lorenz2011} in the price variations. We quantify these socio-economic signals to provide an analytical perspective on the relationship between the Bitcoin exchange rates and the social aspects of its economy. By adopting this perspective, we reveal multiple temporal dependencies leading to the formation of Bitcoin price bubbles.

\paragraph{Four-tiered data}
We use a four-tiered data set (Table~\ref{tab:samplesizes}), composed of records of exchange data, social media activity, search trends, and user adoption of Bitcoin.

\subparagraph{Bitcoin block chain and software client: user base} The Bitcoin block chain is a public ledger containing the full record of all public transactions in the history of the Bitcoin currency \cite{nakamoto2008}. Every node of the Bitcoin network, running a Bitcoin software client, keeps a copy of the block chain. Our analysis of the block chain, as well as of the number of downloads of the software client, yields two approximations for the true number of new Bitcoin users at any time. The number of new Bitcoin users adopting the currency at time $t$ is represented by the variable $U_t$ in Fig.~\ref{fig:timeSeries}b.

\subparagraph{Bitcoin exchange rates.} Bitcoins (BTC) are traded for other currencies at public Internet exchanges. As of December 2013, the oldest public exchange, and the largest to trade BTC for US dollars (USD) and euros (EUR), is Mt. Gox; the largest exchange by BTC volume is BTC China, which trades BTC for Chinese renminbi (CNY) \cite{bitcoincharts2013}. For our analysis, we use trading data from these two exchanges and a third exchange hosted in Europe, BTC-de, including exchange rates in
three currencies USD, EUR, and CNY, using BTC/USD as a historical reference. The trading price is represented by the variable $P_t$ in Fig.~\ref{fig:timeSeries}b.

\subparagraph{Information search.} We measure the interest in obtaining information about Bitcoin through the normalised search volume for the term ``bitcoin'' on the Google search engine. Search volume data have been shown useful to capture the information-gathering stage of the decision process of individuals, leading to insights about volume and volatility \cite{Preis2010,Bordino2012}, as well as financial returns \cite{Preis2013}. Alternatively, Wikipedia usage \cite{Moat2013} can be used as an indicator of information gathering. Both indicators have been shown to lead Bitcoin prices \cite{Kristoufek2013}, motivating the cross-validation of our results with the number of Wikipedia views for the Bitcoin page on the English Wikipedia. The search volume is represented by the variable $S_t$ in Fig.~\ref{fig:timeSeries}b.

\subparagraph{Information sharing.} In our analysis, information search is a private action that nee dnot be shared among individuals, while information sharing is strictly social. Social interaction between individuals can be measured through their level of communication. Previous works applied sentiment analysis on public messages on Twitter (tweets) to predict stock price changes \cite{Bollen2010}, or linguistic patterns in instant messages to predict stock volatility \cite{Saavedra2011}. In addition to sentiment, absolute on-line word-of-mouth levels, such as the total number of tweets or news articles, are useful in predicting price changes \cite{Mao2011}. We measure information sharing, or on-line word-of-mouth communication, through the daily number of Bitcoin-related tweets $B_t$ per million messages in our Twitter feed $T_t$, calculated as $(B_t/T_t) \cdot 10^6$. For additional validation, we compute an alternative by substituting the number of Bitcoin-related tweets with the number of ``reshares'' 
of the messages posted on the oldest, regularly active public Facebook page dedicated to Bitcoin. On-line word of mouth is represented by the variable $W_t$ in Fig.~\ref{fig:timeSeries}b.

\begin{table}
\centering
    \begin{tabular}{|c|c|c|c|}
      \hline
      Start Date & BTC tweets & Total tweets & Wikipedia Views \\
      \hline
      2009-01-09 & $6,827,894$ & $266,306,726,448$ &
      $6,330,676$ \\ \hline
      End Date & Users & Facebook reshares & Client downloads \\ \hline
       2013-10-31 & $4,717,713$ & $2,461$ & $3,817,506$ \\ \hline
    \end{tabular}
    \begin{tabular}{|c|c|c|c|c|}
      \hline
      Market & Currency & Period & BTC Volume & Currency Volume \\
      \hline
      Mt. Gox & USD & 2010-07-17 -- 2013-10-31 & $ 52,273,038.8$ &
      $1,663,743,940.58$ USD \\ \hline
      Mt. Gox & EUR & 2011-09-05 -- 2013-10-31 & $2,748,706.19$ &
      $126,206,290.65$ EUR \\ \hline
      BTC-China & CNY & 2011-06-17 -- 2013-10-31 & $2,062,919.84$ &
      $1,383,924,703.13$ CNY \\ \hline
      BTC-de & EUR & 2011-09-03 -- 2013-10-31 & $958,257.95$ &
      $34,897,155.88$ EUR \\ \hline
    \end{tabular}
      \caption{Sample sizes of the Twitter, Wikipedia, Facebook, Bitcoin network, and SourceForge data sets (top), and BTC exchange markets data sets (bottom).}
  \label{tab:samplesizes}
\end{table}

\section{Materials and Methods}

\paragraph{Internet data sets}
We downloaded the whole Bitcoin block chain, which contains a detailed record of each block, from the website {\footnotesize\url{http://blockexplorer.com}} up to November 5, 2013. We retrieved search volume data from Google Trends on November 5, 2013 {\footnotesize\url{http://www.google.com/trends/explore}}.  We queried for the term ``bitcoin'' for a set of time intervals: first the whole time period (which returned weekly volumes), then a rolling window of two months (returning daily volumes). We combined the results of these queries (see Section S1.1 and Figs.~S1--2 of the ESM) to produce normalised daily search volumes for the whole time period. We retrieved the daily number of views of the Bitcoin page on the English Wikipedia {\footnotesize\url{http://en.wikipedia.org/wiki/Bitcoin}} by using the JSON interface of {\footnotesize\url{http://stats.grok.se}}. We queried each month of the dataset, getting the total amounts of views for the page each day of the queried month. We gathered the relative volume 
of tweets about Bitcoins through Topsy ({\footnotesize\url{http://topsy.com}}) on November 5, 2013, by dividing the number of tweets containing at least one of the following terms: ``BTC'',``\#BTC'', ``bitcoin'', or ``\#bitcoin'' by the total number of tweets for each day of the study period. We extracted the number of ``reshares'' of posted items on the 
oldest (to the best of our knowledge), regularly active public Facebook page dedicated to Bitcoin {\footnotesize\url{http://www.facebook.com/bitcoins}}. We downloaded Bitcoin market data from {\footnotesize\url{http://bitcoincharts.com}}, covering the largest markets for three currencies: Mt. Gox for USD, BTC-China for CNY, and Mt. Gox and BTC-de for EUR. In the main text, we use the Mt. Gox BTC/USD time series as a reference for price when not mentioned otherwise, since it is the largest market by volume throughout the study period \cite{bitcoincharts2013}.

\paragraph{Reconstructing the number of users of the Bitcoin network}
We used two proxies for the number of Bitcoin users: The first is the daily number of downloads of the official Bitcoin software client from the SourceForge platform ({\footnotesize\url{http://sourceforge.net/projects/bitcoin}}). We also analysed the full block chain; the Bitcoin protocol is designed to preserve to a certain extent the anonymity of its users and their activity by identifying them through public keys only \cite{nakamoto2008}. However, heuristics applied on transaction logs make it possible to approximately map sets of public keys to unique users \cite{androulaki2012,reid2013}. We reconstructed the number of users of the Bitcoin network by analysing the complete set of transactions from the block chain data and applying one rule for key aggregation and one heuristic for change identification (these are described in detail in Section S1.2 of the ESM).

\paragraph{Time series stationarity} 
Prior to our analysis, we performed stationarity tests of the time series  $U_t$, $P_t$, $S_t$, and $W_t$, revealing that these variables are integrated of order 1: whilst the time series of levels cannot be assumed to be stationary, the time series of their differences are. 
In the following, for each time series $x(t)$ we calculate the differentiated time series $\Delta x(t) = x(t)-x(t-1)$. 
For each of our four variables, the hypothesis of non-stationarity can be rejected at the 0.01 confidence level, and the hypothesis of stationarity cannot be confidently rejected (Table~S1 of the ESM).

\paragraph{Vector autoregression} 
Since the first differences of our four variables are stationary, a vector autoregression technique (VAR) can reveal the interaction between variables.
We fitted a vector autoregressor of lag 1 with a linear and a seasonal trend, measuring the time-dependent relations between the normalised changes of the four variables of our study. The significance of the relation between variables serves as a multidimensional extension of a Granger (non)-causality test: low p-values allow us to reject the hypothesis that the changes in one variable have no linear relation to the changes of another variable in the previous day. In addition, we calculated the impulse response functions \cite{Lutkepohl2007} of each variable to exogenous shocks on other variables, in the presence of correlated noise. We chose to combine vector autoregression and impulse response functions to account for the multidimensional nature of our analysis and the finite sample size of our data. This provides an advantage in comparison with cross-correlation analysis between pairs of variables \cite{Lutkepohl2007}, which leads to incomplete results, as we report in Section S3 of the ESM.

\paragraph{Fundamental value} It is difficult to calculate an estimate of the fundamental, or intrinsic, value of one bitcoin, which is different to its ``fair'' value \cite{iaralov2013}. However, we argue that the fundamental value of one bitcoin equals at least the cost involved in its production (through mining), and therefore that we can use this cost as a lower-bound estimate of the fundamental value. This definition has the advantage of being independent from any subjective assessment of future returns. This estimate is given by dividing the cumulated mining hash rate in a day by the number of bitcoins mined \cite{blockchaininfo2013}, to obtain the number of SHA-256 hashes needed to mine one bitcoin (this is another way to express the difficulty \cite{nakamoto2008}). We then use an approximation of the power requirements for mining of 0.5W per MHash/s, which is the average efficiency of the most common graphics processing units (GPUs) used to mine bitcoins \cite{bitcoinwiki2013} during our study period 
(mid-2010 to late 2013), and an approximation of electricity costs of \$0.15/KWh, which is an average of US and EU prices \cite{energycosts2013}. This yields our lower bound estimate of the fundamental value of a bitcoin, in \$/BTC.

\section{Results}

\paragraph{Feedback loops between variables} 
We disentangle the feedback cycles in our system by means of a vector autoregressor (VAR) \cite{Toda1994}, which captures time-dependent multidimensional linear relations between the four variables of the analysis, with a lag of one day. In this statistical model, the change in each variable on a given day $\{\Delta U_t,\Delta P_t,\Delta S_t,\Delta W_t\}$ is a linear combination of the changes in all variables $\{\Delta U_{t-1},\Delta P_{t-1},\Delta S_{t-1},\Delta W_{t-1}\}$ on the previous day, including a deterministic, a periodic, and an error term \cite{Whittle1953, Toda1994}. The weight of the change of variable $X_{t-1}$ in the equation describing the change in variable $Y_t$ is denoted $\phi_{X,Y}$. Fig.~\ref{fig:timeSeries}b and Table~\ref{tab:VAR} present a summary of these multivariate relations. The VAR reveals the following
feedback cycles:
\begin{itemize}
\item ``social'' cycle: search volume increases with price ($\phi_{P,S}=0.386$), word of mouth increases with search volume ($\phi_{S,W}=0.243$), and price increases with word of mouth ($\phi_{W,P}=0.1$). Simultaneously accounting for all dependencies between the four variables emphasises the influence of word of mouth on price, revealing a stronger relation than cannot be observed with pairwise correlation analysis (more details in Section S3 of the ESM). The three-way loop between $S_t$, $W_t$ and $P_t$ represents the feedback cycle between social dynamics and price in the Bitcoin economy.
\item ``user adoption'' cycle: search volume increases with price ($\phi_{P,S}=0.386$), the amount of new users increases with search interest ($\phi_{S,U}=0.158$), and price increases with increases in user adoption ($\phi_{U,P}=0.137$). This second three-way loop between $S_t$, $U_t$ and $P_t$ models how the exchange rate of Bitcoin to other currencies depends on the number of users in the Bitcoin economy.
\end{itemize}

\begin{table}[h]
    \begin{tabular}{|l|c|c|c|c|}
       \hline
          & \footnotesize{$\Delta P_{t-1}$} & \footnotesize{$\Delta U_{t-1}$} & \footnotesize{$\Delta W_{t-1}$} & \footnotesize{$\Delta S_{t-1}$} \\
      \hline
\footnotesize{$\Delta P_{t}$} & $\bf{0.153}$ \hfill \footnotesize{$(7.2*10^{-08})$} & $\bf{0.137}$ \hfill \footnotesize{$(5.3*10^{-05})$} & $\bf{0.100} $ \hfill \footnotesize{$(4.4*10^{-04})$} & $\bf{-0.233}$ \hfill \footnotesize{$(2.3*10^{-11})$}\\ \hline
\footnotesize{$\Delta U_{t}$} & $\bf{0.184}$ \hfill \footnotesize{$(1.6*10^{-10})$} & $\bf{-0.143}$ \hfill \footnotesize{$(3.0*10^{-05})$} & $-0.007 $ \hfill \footnotesize{$(7.8*10^{-01})$} & $\bf{0.158}$ \hfill \footnotesize{$(5.8*10^{-06})$}\\ \hline
\footnotesize{$\Delta W_{t}$} & $-0.032$ \hfill \footnotesize{$(2.0*10^{-01})$} & $8.201$ \hfill \footnotesize{$(9.9*10^{-01})$} & $\bf{-0.320} $ \hfill \footnotesize{$(9.4*10^{-34})$} & $\bf{0.243}$ \hfill \footnotesize{$(1.3*10^{-14})$}\\ \hline
\footnotesize{$\Delta S_{t}$} & $\bf{0.386}$ \hfill \footnotesize{$(1.5*10^{-46})$} & $0.016$ \hfill \footnotesize{$(5.9*10^{-01})$} & $-0.013 $ \hfill \footnotesize{$(5.9*10^{-01})$} & $\bf{0.293}$ \hfill \footnotesize{$(5.0*10^{-20})$}\\ \hline
  \end{tabular}
  \caption{Vector autoregression results (weights and p-values) for $P_t$ as Mt. Gox price, $S_t$ as Google search volume, $U_t$ as number of downloads of the Bitcoin client, and $W_t$ as tweet ratio. Results in boldface are significant at the 0.05 level. \label{tab:VAR}}
\end{table}

In addition to these two cycles, we find a negative relation from search to price ($\phi_{S,P}=-0.233$). This is illustrated by a clear dyadic relation between the two variables' extremes: 3 of the 4 largest daily price drops were preceded by the 1st, 4th, and 8th largest increases in Google search volume the day before. Table~S2 of the ESM presents the results of the same VAR with non-normalised time series; in this way, relations between variables can effectively be quantified (e.g., an increase of 10'000 client downloads leads to an increase of \$3.80 in price)

These two feedback cycles and the negative role of search in price are consistent when fitting vector autoregressors using the number of users in the network instead of client downloads as user signal $U_t$, Wikipedia views instead of Google search volume as search signal $S_t$, and Facebook Bitcoin page reshares instead of Bitcoin-related tweets as $W_t$ (Table~S3). All these results are also consistent when using other currency pairs (BTC/EUR and BTC/CNY) and data from other exchange platforms (BTC-China and BTC-de) for the analysis (Table~S4).

\begin{figure}[ht!]
  \centering
\includegraphics[width=\textwidth]{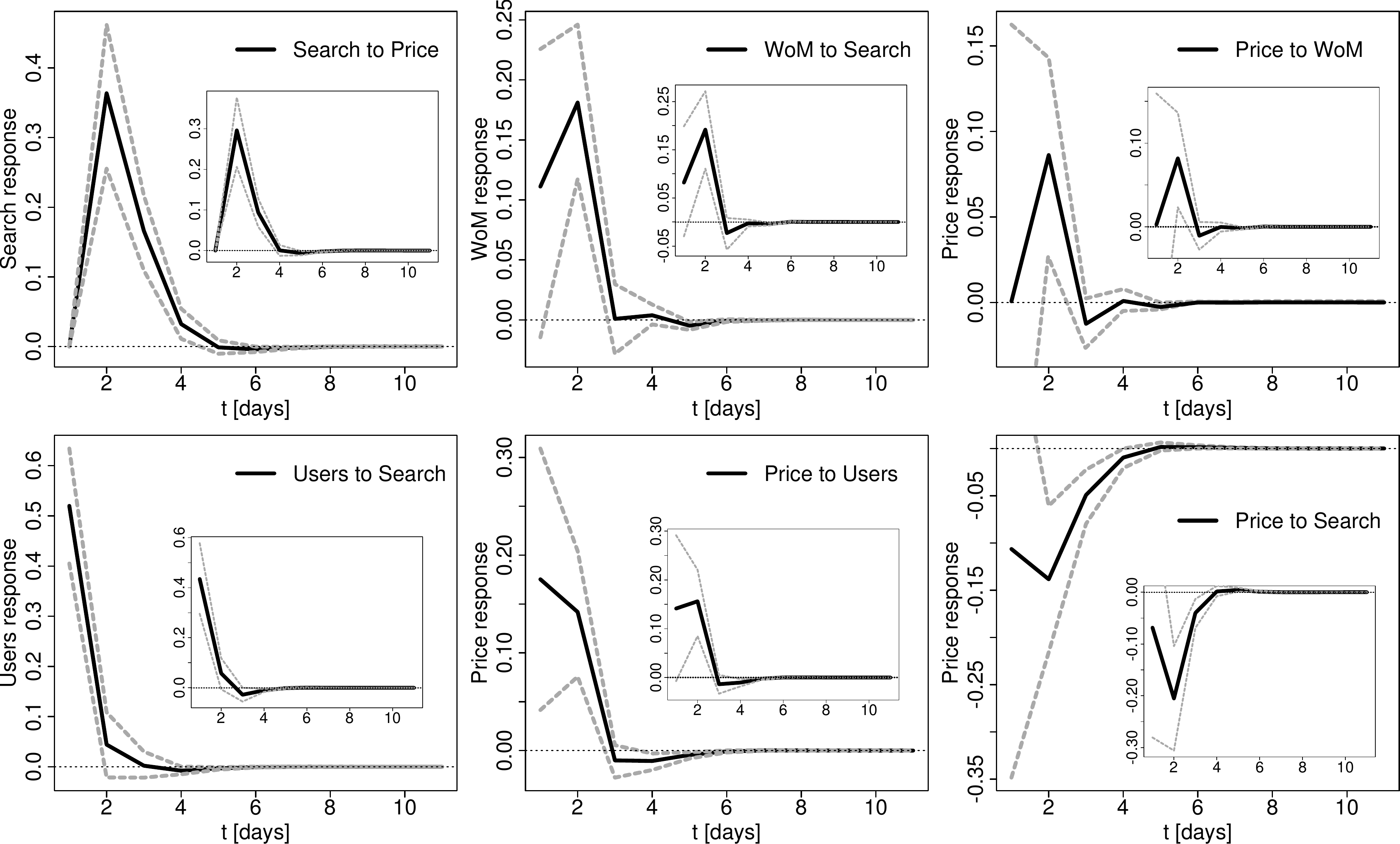}
\caption{Impulse response function point estimates and 95\% confidence intervals for the feedback loops between variables in our model, plus the negative relation between search and price. The inset shows results using Wikipedia views instead of Google search volumes as an estimator for $S_t$.}
\label{fig:IRF}
\end{figure}

To further verify the connection between the variables in the social and user adoption cycles, we calculated the impulse response functions of the vector autoregression results shown in Table~\ref{tab:VAR}. 
The impulse response function estimates the propagation
of a shock of one standard deviation on a variable to the other variables. To control for our finite sample size, we performed 10000 boostrapped estimations, correcting for correlated noise with the standard HAC method \cite{Zeileis2004}. The results of the estimation for the response of variables and their 95\% confidence intervals are shown in Fig.~\ref{fig:IRF}.
All the pairwise relations of the feedback cycles illustrated in Fig.~\ref{fig:timeSeries} are significant when analysed through impulse response functions. Search levels experience significant increases two to four days after price increases, and both word of mouth and number of users increase one to two days after strong increases in search volume. Price increases as a result of shocks on user adoption and word of mouth, and the negative influence of search in price is also evident. In contrast with findings in previous time periods \cite{Kristoufek2013}, these response functions are similar for both Google search trends and Wikipedia page views. To further evaluate the significance of these responses, we measured the response estimate between each pair of signals for time increments of 1 and 2 days, testing for a significance of 97.5\%, to achieve a 95\% confidence level after Bonferroni correction. All the relations in the feedback cycles have at least one significant change, including the negative 
effect of search trends to price (more details in Table~S7 of the ESM).

\begin{figure}[ht!]
  \centering
  \includegraphics[width=0.95\textwidth]{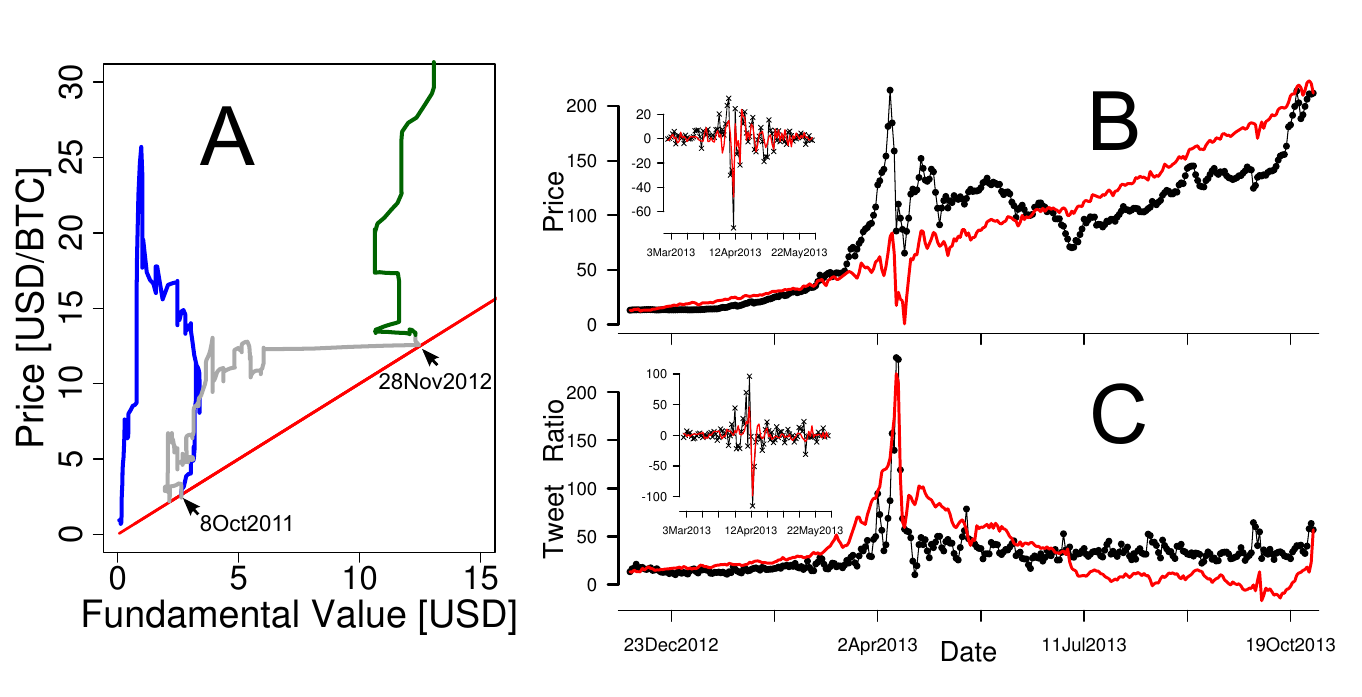} \caption{{\bf
      A.}  Trajectory of weighted mean price and mean fundamental
    value over a rolling window of width of one week (one point per
    day, connected in chronological order) between March 1st, 2011 and
    March 1st, 2013. The red line denoting strict equality between
    price and fundamental value is hit on two occasions (around
    October 18th, 2011, and November 28th, 2012), which delimit the
    characteristic periods of our analysis.  The blue line shows values
    before the first hitting event, the green line shows the values after
    the last hitting event, and the grey line shows the period in between.
    {\bf B, C.} Time series of the cumulative price and tweet ratio
    (black dots), and cumulative estimated values by the model for the
    period since November 30, 2012. Insets: time series of estimates
    and empirical changes of price and Tweet ratio for the period
    between March 1 and May 30, 2013.}
  \label{fig:fits}
\end{figure}

\paragraph{Reproducing price and social dynamics.} The cycles
presented above provide an explanation for the generation of bubbles
in the Bitcoin economy.  Recent findings indicate that the driving
forces behind Bitcoin prices changed since its invention
\cite{Kristoufek2014}, motivating our decomposition of the study
period into characteristic time windows, each of which corresponds to
a distinct bubble.  We do this by estimating a lower bound for the
fundamental value of Bitcoin: we approximate the energy cost of
producing one bitcoin, which is derived directly from the Bitcoin
difficulty \cite{nakamoto2008} (see Materials and Methods). Throughout
our study period, the price stayed almost always above the fundamental
value (Fig.~\ref{fig:fits}a: the trajectory of the weekly weighted
mean price is almost exclusively on the left of the price/fundamental
equality line). The trade of bitcoins at a much higher price indicates
the possible presence of a bubble \cite{sornette2009}, and the events
at which the market price starts diverging from the fundamental value
mark the beginning of bubbles. We identify two such events: one around
October 18th, 2011, which marked the end of the 2011 bubble
\cite{theeconomist2011}, and another one around November 28th, 2012,
which is the date at which mining rewards halved (``Halving Day''
\cite{btcfoundation2012}), suddenly raising the fundamental
value. During the study period the price never dropped significantly
below the fundamental value, which validates our estimate for it: a
drop of price below the fundamental value would introduce a paradox in
which the maintenance of the public ledger by the miners is no more
profitable for them. We thus define three characteristic periods, or
bubbles (Fig.~\ref{fig:timeSeries}b): before the first event (P1),
between the first and the second events (P2), and after the second
event (P3).  Running independent VAR analyses on the three periods
shows that the two feedback cycles exist in the system when
considering the three periods together or P3 on its own, but not if we
only take into account data until November 28th, 2012 (Table~S5 of the
ESM). In the following, we focus on the last period to evaluate the
quality of the VAR technique in reproducing changes in the four
variables in the last bubble.

The results shown in Table \ref{tab:VAR}, with a lag of one day, allow us to interpret our results in terms of feedback cycles. Nevertheless, an extended VAR with a longer lag can have larger explanatory power, despite lacking a straightforward interpretation \cite{Sims1990, Toda1994}. By assessing the quality of the model through the Schwarz criterion \cite{schwarz1978}, we find the optimal explanatory power corresponds to a lag of four days (Fig.~S4 of the ESM). In this best fit, the changes in each variable are calculated as linear combinations of the changes of all variables up to four days before. We use this optimised model to estimate daily changes in our four variables $\{\Delta U_t,\Delta P_t,\Delta S_t,\Delta W_t\}$ based on the empirical data. Adding the daily changes yields a step-wise reconstruction of the time series of each variable. Fig.~\ref{fig:fits}b,c show the overlaid times series of estimated and empirical price and word-of-mouth levels. While this method fits daily changes in the 
variables, we can assess the long-term quality of the VAR by correlating their cumulative time series with the reconstructed values. We find a positive significant Pearson's correlation coefficient $\rho$ between the observed and estimated time series for all four variables, with the price comparison yielding $\rho=0.8422$, and the word-of-mouth comparison $\rho=0.6155$. Additionally, reconstructed time series for the number of users ($\rho=0.2261$) and search volume ($\rho=0.6838$) are also accurate ($p<10^{-10}$ for all correlations. Residuals from the fit of the changes are approximately normally distributed (Fig.~S5 of the ESM), which means that within period P3 there are no structural deficiencies in our model. Finally, the VAR model correctly identifies the sign of all of the 10 largest daily price increases, and 9 of the 10 largest price drops during P3 (Table~S6 of the ESM).

\section{Discussion}
Due do the decentralised character of the Bitcoin currency, the dynamics of its economy largely depend on the behaviour of its users, who (i) mine new bitcoins and maintain the block chain, and (ii) influence the exchange rate by trading bitcoins to and from other currencies; these interactions between users form the social backbone of Bitcoin. In this paper, we used the digital traces of user activity in the Bitcoin economy to disentangle the relationships between its different variables: user base, information search, information sharing, and price. While each of these four socio-economic signals can be studied independently, the novelty of our approach lies in the combination of all of them into a robust analytical approach. Together, these signals provide a precise picture of Bitcoin's growth over the three-year period we studied, from the first time bitcoins were publicly traded in the middle of 2010 through the successive price surges until the end of 2013. These digital activity traces \cite{Lazer2009}
 are a unique source of information to quantitatively describe the dynamics of large socio-economic systems \cite{Preis2010,Preis2013}; our analysis combines for the first time social, economic, and technological factors into a unified perspective on one such system.

This combined analysis reveals two positive feedback loops: a reinforcement cycle between search volume, word of mouth, and price (social cycle), and a second cycle between search volume, number of new users, and price (user adoption cycle). The social cycle provides evidence for interindividual influence in the decision to buy bitcoins. It translates as follows: Bitcoin's growing popularity leads to higher search volumes, which in turn result in increased social media activity on the topic of Bitcoin. More interest encourages the purchase of bitcoins by individual users, driving the prices up, which eventually feeds back on the search volumes. We hypothesise that the temporal correlation between price and search volume is mediated by the media reporting on price increases, thereby driving user curiosity and triggering their search activity. The second feedback loop we identify is the user adoption cycle, which complements the social cycle: new Bitcoin users download the client and join the transaction 
network after acquiring information about the technology. This growth in the user base translates to a price increase, as the number of bitcoins available for trade does not depend on demand, but rather grows linearly with time. This is a direct consequence of the deflationary nature of Bitcoin as a currency. Another important result of the VAR is the negative weight of search on price. This marks sporadic connections between large price drops and the spikes in search volume that preceded them. In other words, user search activity responds faster to negative events, such as a security breach in a Bitcoin exchange, than price. In this regard, search spikes are early indicators of price drops.

We showed the robustness of our vector autoregression analysis, computing impulse response functions of the key couplings between variables in the feedback loops.
Furthermore, we validated that our findings are not a construct of the particularities of Mt. Gox as an on-line exchange, reproducing our results in three different currency markets (USD, EUR, and CNY). Our analysis of behavioural signals is also robust to the choice of data sources, as our results also appear when using
alternative sources for search, word of mouth, and user adoption. In the case of user adoption, we found that the number of new users detected after processing the block chain yields information about price changes, which is not observed when using raw Bitcoin addresses without pre-processing (see Table S3 of the ESM). This indicates that identifying users from addresses is important to correctly characterise the Bitcoin ecosystem, which has been overlooked in previous studies \cite{Kondor2014}.

We introduced a lower-bound estimate of Bitcoin's fundamental value, based on the cost of the energy involved in mining. This allowed us to identify three characteristic time periods; by studying the three periods independently, we found that the two feedback cycles are rooted in the last period of the data (P3, end of 2012 until end of 2013), which may either denote that the appearance of these cycles is recent, or presumably that the larger quantity of data from the last period yields a higher signal-to-noise ratio, thus helping the characterisation of the cycles. This second hypothesis would also explain the more nuanced findings obtained by our analysis compared to earlier work on earlier data sets \cite{Kristoufek2013}.

We validated the results of the vector autoregression technique in this last period by comparing the output of the best VAR fit to the empirical time series of our four variables. It is important to note that the VAR provides a fit of the daily changes, but in this validation step we compare the cumulative levels. If the VAR errors were temporally correlated, this would create structural divergence in the VAR estimates of the levels. However, this is not what we find: the levels produced by the VAR are significantly correlated with the empirical values, and the residuals of the estimates are normally distributed (Fig.~S5 of the ESM). We also find that the dynamics of both qualitatively match (Fig.~\ref{fig:fits}), and that the VAR correctly classifies the sign of the largest price variations (Table~S6 of the ESM). The statistical technique we used in this paper thus proves to be a robust way of identifying the coupled dynamics of the socio-economic variables we study. It also produces accurate estimates of 
the future levels of any variable (including price and word of mouth) based on the past history of the system.

The two positive feedback loops we identified imply a constant increase of the price, which should drive up the Bitcoin exchange rate unsustainably. This provides an explanation for the successive periods of almost uninterrupted growth observed from mid-2010 to June 2011, and early 2012 to April 2013. It does not however explain the sudden drops observed at the end of these periods. These crashes can be attributed to external stimuli, such as the attack on the Mt. Gox platform of June 2011. The relationship we found between price drops and preceding search spikes, while it does not explain the occurrence of the crashes, is consistent with past studies linking search volumes to price fluctuations in financial markets \cite{Bordino2012,Preis2013}. While previous work indicated that news sources do not constitute good predictors of future price changes \cite{cutler1989}, our analysis suggests that the successive price surges in the Bitcoin economy are largely due to its growing public attention.
Initially, bitcoins had a negligible exchange value and were only known by a small community of technically experienced users.
Our analysis suggests that the successive waves of growth of the Bitcoin economy were driven by corresponding waves of new users from public circles gradually opening to the currency.
The growing user base created more exchange opportunities for Bitcoin which, at the time of writing, is accepted by a wide range of businesses, from hosting services to retail stores and illegal markets. As of December 2013, the computational power of the Bitcoin mining network sits at about $7\cdot10^{19}$ floating-point operations per second -- roughly 300 times the combined power of the top 500 supercomputers \cite{comppower2013}. This striking figure illustrates the increasing importance of Bitcoin in the technological, social, and economic landscapes, and motivates the design of policies to regulate Bitcoin usage and exchange. For such regulations to be effective, policy makers require an empirical understanding of the dynamics of Bitcoin adoption and trade. The digital traces left by the millions of users of the Bitcoin network, exchange markets, on-line social networks, and search engines allowed us to systematically describe the dynamics of Bitcoin adoption. Our analysis of collective socio-economic 
signals can be applied beyond the study of cryptocurrencies, to understand other phenomena for which large amounts of data are available, such as adoption behaviour in on-line communities \cite{onnela2010,Garcia2013}.

\section*{Acknowledgements}
DG's work was funded by the Swiss National Science Foundation (CR21I1\_146499\/1). We are grateful to Frank Schweitzer for supporting this project.

\newpage
%

\begin{thebibliography}{10}

\bibitem{nakamoto2008}
Nakamoto S.
\newblock {Bitcoin: A peer-to-peer electronic cash system}.
\newblock {Bitcoin Foundation}; 2008.

\bibitem{hadas2013}
Hadas E. {A prediction: Bitcoin is doomed to fail}; 2013.
\newblock Accessed: 2013-12-01.
\newblock The New York Times.
  \url{http://dealbook.nytimes.com/2013/11/27/a-prediction-bitcoin-is-doomed-t%
o-fail}.

\bibitem{blackstone2014}
{The Blackstone Group L P}. {Byron Wien Announces Predictions for Ten
  Surprises for 2014}; 2014.
\newblock Accessed: 2013-12-07.
\newblock Press release.
  \url{http://www.blackstone.com/news-views/press-releases/details/byron-wien-%
announces-predictions-for-ten-surprises-for-2014}.

\bibitem{fontevecchia2013}
Fontevecchia A. {Winklevoss Twins Say Bitcoin Market To Hit \$400B, Urge
  Regulators Not To Push Innovation To China}; 2013.
\newblock Accessed: 2013-12-01.
\newblock Forbes.

\bibitem{blockchaininfo2013}
{Data from blockchain.info}; 2013.
\newblock Accessed: 2013-12-01.
\newblock \url{https://blockchain.info/charts}.

\bibitem{googletrends2013}
{Data from Google Trends}; 2013.
\newblock Accessed: 2013-12-01.
\newblock \url{http://www.google.com/trends}.

\bibitem{Fama1969}
Fama EF, Fisher L, Jensen MC, Roll R.
\newblock The Adjustment of Stock Prices to New Information.
\newblock International Economic Review. 1969;10(1):pp. 1--21.
\newblock Available from: \url{http://www.jstor.org/stable/2525569}.

\bibitem{grossman1976}
Grossman SJ, Stiglitz JE.
\newblock Information and competitive price systems.
\newblock The American Economic Review. 1976;66(2):246--253.

\bibitem{bikhchandani1992}
Bikhchandani S, Hirshleifer D, Welch I.
\newblock {A Theory of Fads, Fashion, Custom, and Cultural Change as
  Informational Cascades}.
\newblock The Journal of Political Economy. 1992;100(5):992--1026.
\newblock Available from: \url{http://dx.doi.org/10.2307/2138632}.

\bibitem{Saavedra2011}
Saavedra S, Duch J, Uzzi B.
\newblock {Tracking traders' understanding of the market using e-communication
  data.}
\newblock PLOS ONE. 2011;6(10):e26705.

\bibitem{lorenz2011}
Lorenz J, Rauhut H, Schweitzer F, Helbing D.
\newblock How social influence can undermine the wisdom of crowd effect.
\newblock {Proceedings of the National Academy of Sciences}.
  2011;108(22):9020--9025.

\bibitem{bitcoincharts2013}
{Data from Bitcoin Charts}; 2013.
\newblock Accessed: 2013-12-07.
\newblock \url{http://bitcoincharts.com/markets}.

\bibitem{Preis2010}
Preis T, Reith D, Stanley HE.
\newblock {Complex dynamics of our economic life on different scales: insights
  from search engine query data.}
\newblock Philosophical transactions Series A, Mathematical, physical, and
  engineering sciences. 2010;368(1933):5707--19.

\bibitem{Bordino2012}
Bordino I, Battiston S, Caldarelli G, Cristelli M, Ukkonen A, Weber I.
\newblock {Web search queries can predict stock market volumes.}
\newblock PLOS ONE. 2012;7(7):e40014.

\bibitem{Preis2013}
Preis T, Moat HS, Stanley HE.
\newblock {Quantifying trading behavior in financial markets using Google
  Trends}.
\newblock Scientific Reports. 2013;3:1684.

\bibitem{Moat2013}
Moat HS, Curme C, Avakian A, Kenett DY, Stanley HE, Preis T.
\newblock {Quantifying Wikipedia usage patterns before stock market moves}.
\newblock Scientific reports. 2013;3:1801.

\bibitem{Kristoufek2013}
Kristoufek L.
\newblock {BitCoin meets Google Trends and Wikipedia: Quantifying the relationship between phenomena of the Internet era}.
\newblock Scientific reports. 2013;3:3415.

\bibitem{Bollen2010}
Bollen J, Mao H, Zeng Xj.
\newblock {Twitter mood predicts the stock market}.
\newblock Journal of Computational Science. 2011;2:1--8.

\bibitem{Mao2011}
Mao H, Counts S, Bollen J.
\newblock {Predicting Financial Markets: Comparing Survey, News, Twitter and
  Search Engine Data}. 2011;p.~10.
\newblock Available from: \url{http://arxiv.org/abs/1112.1051}.

\bibitem{androulaki2012}
Androulaki E, Karame G, Roeschlin M, Scherer T, Capkun S.
\newblock {Evaluating User Privacy in Bitcoin} [IACR Cryptology ePrint
  Archive]. 2012;p. 596.

\bibitem{reid2013}
Reid F, Harrigan M.
\newblock {An analysis of Anonymity in the Bitcoin System}.
\newblock In: Security and Privacy in Social Networks. Springer; 2013. p.
  197--223.

\bibitem{Lutkepohl2007}
L\"utkepohl H.
\newblock {New introduction to multiple time series analysis}.
\newblock Springer, 2007.

\bibitem{iaralov2013}
Woo D, Gordon I, Iaralov V.
\newblock {Bitcoin: a first assessment}.
\newblock {Bank of America Merrill Lynch}; 2013.

\bibitem{bitcoinwiki2013}
{Data from the Bitcoin wiki}; 2013.
\newblock Accessed: 2013-12-07.
\newblock \url{https://en.bitcoin.it/wiki/Mining_hardware_comparison}.

\bibitem{energycosts2013}
{Data from the U.S. Energy Information Administration and the European
  Commission}; 2013.
\newblock Accessed: 2013-12-07.
\newblock
  \url{http://www.eia.gov/electricity/monthly/epm_table_grapher.cfm?t=epmt_5_6%
_a} and
  \url{http://epp.eurostat.ec.europa.eu/statistics_explained/index.php/Energy_%
price_statistics}.

\bibitem{Whittle1953}
Whittle P.
\newblock The Analysis of Multiple Stationary Time Series.
\newblock Journal of the Royal Statistical Society Series B (Methodological).
  1953;15(1):pp. 125--139.

\bibitem{Toda1994}
Toda HY, Phillips PCB.
\newblock Vector autoregression and causality: a theoretical overview and
  simulation study.
\newblock Econometric Reviews. 1994;13(2):259--285.

\bibitem{Zeileis2004}
Zeileis A.
\newblock Econometric Computing with HC and HAC Covariance Matrix Estimators.
\newblock Journal of Statistical Software. 2004;11(10):1--17.

\bibitem{Kristoufek2014}
Kristoufek L.
\newblock {What are the main drivers of the Bitcoin price? Evidence from wavelet coherence analysis}.
\newblock 2014, available from: \url{http://arXiv:1406.0268}

\bibitem{sornette2009}
Sornette D.
\newblock Why stock markets crash: critical events in complex financial
  systems.
\newblock Princeton University Press; 2009.

\bibitem{theeconomist2011}
F C. {The bursting of the Bitcoin bubble}; 2011.
\newblock Accessed: 2013-12-08.
\newblock The Economist.
  \url{http://www.economist.com/blogs/babbage/2011/10/virtual-currencies}.

\bibitem{btcfoundation2012}
{Bitcoin Foundation.} {Bitcoin Community Celebrates "Halving Day"}; 2012.
\newblock Accessed: 2013-12-08.
\newblock Press release.
  \url{http://www.prlog.org/12032578-bitcoin-community-celebrates-halving-day.%
html}.

\bibitem{Sims1990}
Sims CA.
\newblock Macroeconomics and reality.
\newblock Modelling Economic Series Clarendon Press, Oxford. 1990;.

\bibitem{schwarz1978}
Schwarz G.
\newblock Estimating the dimension of a model.
\newblock {The Annals of Statistics}. 1978;6(2):461--464.

\bibitem{Lazer2009}
Lazer D, Pentland A, Adamic LA, Aral S, Barabasi AL, Brewer D, et~al.
\newblock {Computational social science.}
\newblock Science. 2009 Feb;323(5915):721--723.

\bibitem{Kondor2014}
Kondor D, P\'{o}sfai M, and Csabai I and Vattay G.
\newblock {Do the Rich Get Richer? An Empirical Analysis of the Bitcoin Transaction Network.}
\newblock PLoS ONE. 2014 Feb;9(2)e86197.

\bibitem{cutler1989}
Cutler DM, Poterba JM, Summers LH.
\newblock {What Moves Stock Prices?}
\newblock National Bureau of Economic Research; 1989. 2538.
\newblock Available from: \url{http://www.nber.org/papers/w2538}.

\bibitem{comppower2013}
Data from www.bitcoinwatch.com and www.top500.org; 2013.
\newblock Accessed: 2013-12-02.
\newblock \url{http://www.bitcoinwatch.com}, \url{http://www.top500.org}.

\bibitem{onnela2010}
Onnela JP, Reed-Tsochas F.
\newblock Spontaneous emergence of social influence in online systems.
\newblock Proceedings of the National Academy of Sciences.
  2010;107(43):18375--18380.

\bibitem{Garcia2013}
Garcia D, Mavrodiev P, Schweitzer F.
\newblock Social Resilience in Online Communities: The Autopsy of Friendster.
\newblock In: Proceedings of the First ACM Conference on Online Social
  Networks. COSN '13; 2013. p. 39--50.

\end{thebibliography}
%

\newpage



\section*{Supplementary material}

# The digital traces of bubbles: feedback cycles between socio-economic signals in the Bitcoin economy

## David Garcia, Claudio J. Tessone, Pavlin Mavrodiev, Nicolas Perony

# S1 Electronic Supplementary Material

## S1.1 Reconstructing Google search time series

We retrieved Google trends data for the term "bitcoin" on November 5th, 2013, using the csv export function[1]. This service provides weekly rescaled search volumes since 2004, but daily search volumes can only be downloaded for time intervals up to three months. We retrieved the time series of daily search volumes for the period since the beginning of 2010 as follows:

- First, we do one query for the weekly rescaled volumes since Jan 1st 2010, represented as the blue dots in Fig. S1.

- Second, we perform a set of queries for a rolling time window of two months, returning daily volumes. For each week in the whole time period, we have the rescaled daily volumes, shown as the red inset in Fig. S1.

- Third, we add the daily resolution to the weekly volumes by rescaling the intra-week volumes of the second step, approximating this way the whole time series of daily search volumes.

To validate the correctness of this reconstruction method, we applied it to simulated time series of Brownian motion, white and pink noise. For each simulated time series, we create a weekly volume version, and a set of intra-week volumes as we got from Google trends. We use that data to reconstruct the simulated time series, and evaluate the error introduced by the reconstruction method comparing the initial and the reconstructed time series. Fig. S2 shows a simulation of Brownian motion, and its reconstructed values. The inset shows the scatter plot of both simulated and reconstructed values, showing that the reconstruction method provides a very precise, rescaled approximation of the original time series. This way, less than 0.2% of the variance was lost for Brownian motion, and less than 0.05% was lost for white and pink noise.

---

[1]http://www.google.com/trends/explore

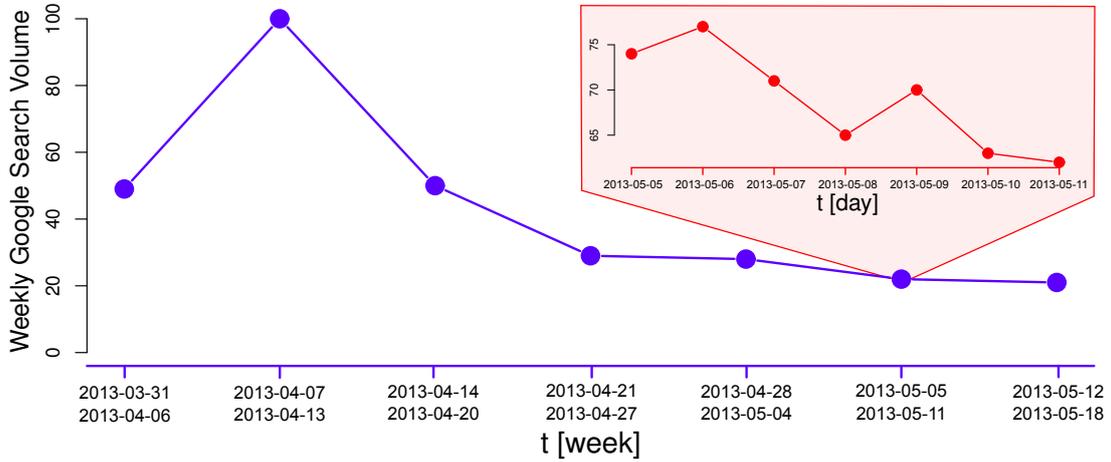

Figure S1: Example of Google Search Volume data. Search volume from Google trends can be extracted in two formats: weekly normalised volume (blue), and daily normalised volume (red, inset). Weekly volume data spans the whole analysis period, daily data can be retrieved for periods up to three months. Our reconstruction method combines these two sources to produce a daily time series for the whole period.

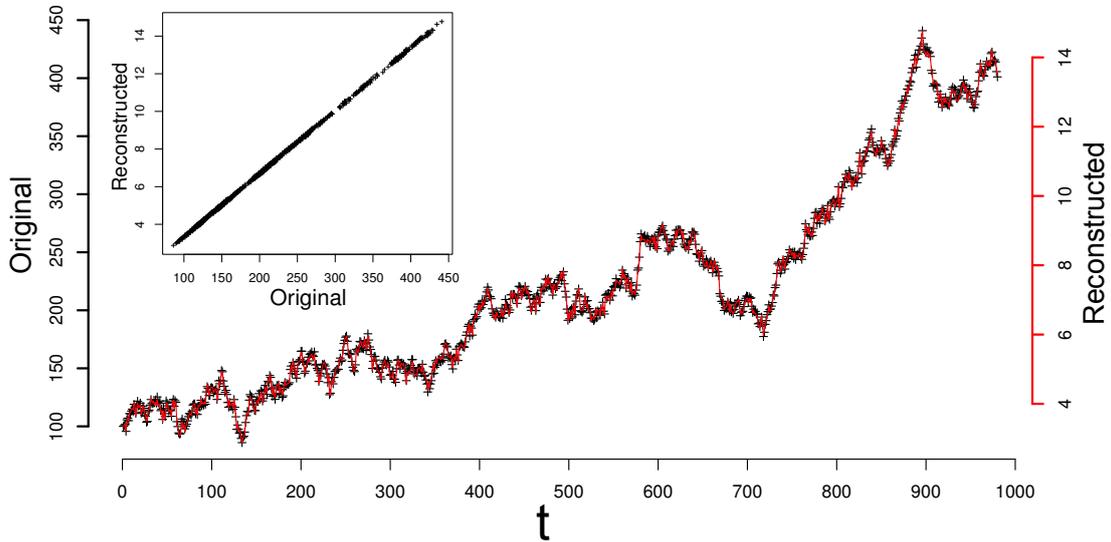

Figure S2: Results of reconstruction method. A simulated time series of Brownian motion (black dots), and its reconstructed version (red lines). Inset: scatter plot of simulated daily values and reconstructed ones. The reconstruction method loses less than 0.2% of the variance of the original time series. Additional tests with white and pink noise lose less than 0.05% of the variance.

## S1.2 Reconstructing the number of Bitcoin users from the block chain dataset

A Bitcoin address is a unique 27-34 alphanumeric identifier of the following form –
`31uEbMgunupShBVTewXjtqbBv5MndwfXhb`. It represents a possible destination for a Bitcoin payment. A Bitcoin user can have multiple addresses. To extract the addresses used for transacting in the Bitcoin network and to later identify unique users, we analysed the Bitcoin block chain up to November 2013. The block chain is a growing historical "book of records" containing full information about all transactions that take place in the network. Every Bitcoin client keeps a complete copy of the block chain, stored locally in the form of raw binary data. Once a user A decides to send a certain amount of BTC to user B, the client creates a transaction is listing (i) the desired amount to transfer, (ii) all of user A's addresses from which BTCs, adding up at least to the desired amount, are to be collected (the input field) and (iii) the addresses to which the desired amount should be sent (the output field). This transaction is then broadcasted to the network, and eventually written into the last block of the block chain. Since the block chain contains information about addresses only, we use the following two rules to match some of these addresses to individual users:

1. **Merging multiple input addresses.** As mentioned above, the input field of a transaction contains all addresses whose total BTC content must be combined for the transfer. To access the funds in a Bitcoin address, a user must possess a secret number, known as a "private key", which implies that every Bitcoin address has a unique owner. It follows, then, that all input addresses must belong to the same user – the one who knows the private key[2]. Hence, for each transaction we merge the input addresses and assign them to a unique user. Moreover, if a transaction has an input address which has already been assigned then the remaining input addresses are also assigned to the same user.

2. **Identifying change addresses** Due to the design of the Bitcoin protocol, the content of a Bitcoin address must be fully spent, once this address has been chosen to be an input for a transaction. Consequently, the total amount of BTC contained in the transaction input must be fully spent. Often in practice, the total funds in the inputs are in excess of the desired transfer amount. In such case, the remaining Bitcoins must be returned to the sender as "change". In the vast majority of current use, the Bitcoin client accomplishes this by creating internally a new valid Bitcoin address, owned by the sender, to which the change is to be transferred. The funds collected in this change address can be reclaimed by the Bitcoin client for the next transaction of the sender. Moreover, the process is invisible to the user – the change address is not known, unless the user manually inspects the output

---

[2]It is highly unlikely that multiple users control the different input addresses, as this implies that they have shared the private keys among them

field of the transaction in the block chain. Therefore, it is unlikely that the user will provide this address as recipient for a future transaction, which implies that the change address appears only once in the output field of a transaction. This is our second rule – we scan the output field of each transaction and look for an address that has only been used once as output in the block chain. If there is only one such address in a transaction, we assign it to the sender. In case of multiple addresses meeting this criterion, we conservatively do not label a change address for this transaction.

## S2   Time series stationarity

| $X$ | Data | ADF | KPSS | $A(1)$ | $\Delta X$ | ADF | KPSS | $A(1)$ |
|---|---|---|---|---|---|---|---|---|
| $W$ | Twitter | $< 10^{-30}$ | $< 0.01$ | 0.816 | $\Delta W$ | $< 10^{-57}$ | $> 0.1$ | $-0.299$ |
| $W$ | Facebook | $< 10^{-57}$ | $< 0.01$ | 0.302 | $\Delta W$ | $< 10^{-57}$ | $> 0.1$ | $-0.372$ |
| $S$ | Google | $< 10^{-14}$ | $< 0.01$ | 0.999 | $\Delta S$ | $< 10^{-57}$ | $> 0.1$ | 0.277 |
| $S$ | Wikipedia | $< 10^{-20}$ | $< 0.01$ | 0.929 | $\Delta S$ | $< 10^{-57}$ | $> 0.1$ | 0.145 |
| $U$ | Client Downloads | $< 10^{-10}$ | $< 0.01$ | 0.990 | $\Delta U$ | $< 10^{-57}$ | $> 0.1$ | $-0.036$ |
| $U$ | Blockchain Users | $< 10^{-9}$ | $< 0.01$ | 0.999 | $\Delta U$ | $< 10^{-57}$ | $> 0.1$ | $-0.033$ |
| $U$ | Addresses | $< 10^{-12}$ | $< 0.01$ | 0.999 | $\Delta U$ | $< 10^{-57}$ | $> 0.1$ | $-0.164$ |
| $P$ | MtGox | 0.92 | $< 0.01$ | 0.999 | $\Delta P$ | $< 10^{-57}$ | 0.0822 | 0.178 |
| $P$ | BTCChina | 0.24 | $< 0.01$ | 0.998 | $\Delta P$ | $< 10^{-57}$ | $> 0.1$ | 0.326 |
| $P$ | MtGox EUR | 0.38 | $< 0.01$ | 0.998 | $\Delta P$ | $< 10^{-49}$ | $> 0.1$ | 0.216 |
| $P$ | BTC-de | 0.25 | $< 0.01$ | 0.998 | $\Delta P$ | $< 10^{-57}$ | $> 0.1$ | 0.23 |

Table S1: Results of stationarity tests for each variable $X$ and their first difference $\Delta X$.

## S3   Lagged correlation analysis

We study the feedback dynamics between these four variables by means of a lagged correlation analysis. Fig. S3 shows the result of lagged Pearson's cross-correlation tests ($\rho(X(t), Y(t + \delta t))$) between all pairs of our four variables. We observe that changes in the exchange rate of USD to BTC on Mt. Gox (hereafter price) lead to corresponding changes in search volume, with $\rho$ peaking at one day (fast response). However, increases in search volume lead to decreases in price after a few days. Increases in price lead to an increased number of client downloads after 1-2 days. Search volume precedes word-of-mouth levels (tweet ratio) by one day, showing that information search precedes information sharing. There are faint positive correlations between

client downloads and word of mouth, as well as between word of mouth and price, but the lagged correlation analysis is not sufficient to conclude dependencies between these two variables, which motivates our vector autoregression analysis. All the relations reported here are consistent when using Wikipedia views instead of Google searches as the variable for $S_t$. These two ways of measuring search interest are strongly correlated at lag $\delta = 0$. While such an analysis gives useful insights into the dynamics of the strict pairwise correlations between the variables of our study, it does not account for coupled correlations between multiple variables, for example between search and users on price. Our VAR analysis takes into account the multidimensional nature of our analysis, and reveals stronger relations that were not observable in this pairwise analysis.

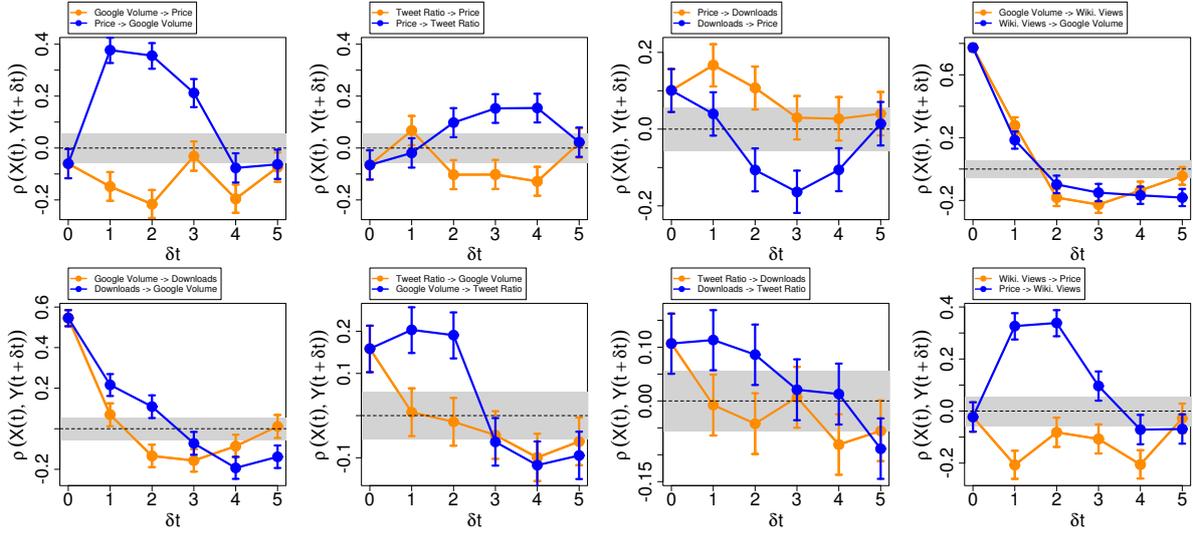

Figure S3: Cross-correlation function between pairs of variables in our analysis. Dots indicate Pearson's correlation coefficient between lagged variables, error bars show 95% confidence intervals. The greyed area around 0 shows the average confidence interval of the correlation for 1000 permutations of the empirical data.

## S4 VAR Results without normalisation

|  | $\Delta P_{t-1}$ | $\Delta U_{t-1}$ | $\Delta W_{t-1}$ | $\Delta S_{t-1}$ |
|---|---|---|---|---|
| $\Delta P_t$ | **0.153** $(7.2*10^{-08})$ | **0.00038** $(5.3*10^{-05})$ | **0.027** $(4.4*10^{-04})$ | **−1.867** $(2.3*10^{-11})$ |
| $\Delta U_t$ | **67.166** $(1.6*10^{-10})$ | **−0.143** $(3.0*10^{-05})$ | −0.790 $(7.8*10^{-01})$ | **462.059** $(5.8*10^{-06})$ |
| $\Delta W_t$ | −0.117 $(2.0*10^{-01})$ | 8.186 $(9.9*10^{-01})$ | **−0.320** $(9.4*10^{-34})$ | **7.070** $(1.3*10^{-14})$ |
| $\Delta S_t$ | **0.048** $(1.5*10^{-46})$ | 5.700 $(5.9*10^{-01})$ | −0.000 $(5.9*10^{-01})$ | **0.293** $(5.0*10^{-20})$ |

Table S2: Vector Auto-Regression results for $P_t$ as price on Mt. Gox, $W_t$ as tweet ratio, $S_t$ as Google search volume, and $U_t$ as number of client downloads, without renormalisation.

## S5  VAR results with variable replacements

|          | $\Delta P_{t-1}$ | $\Delta U_{t-1}$ | $\Delta W_{t-1}$ | $\Delta S_{t-1}$ |
|----------|------------------|------------------|------------------|------------------|
| $\Delta P_t$ | **0.158**  $(1.3*10^{-08})$ | **0.148**  $(3.1*10^{-06})$ | **0.094**  $(7.4*10^{-04})$ | **$-$0.291** $(2.5*10^{-19})$ |
| $\Delta U_t$ | **0.176**  $(7.9*10^{-10})$ | $-0.120$ $(2.1*10^{-04})$ | $-0.000$  $(9.8*10^{-01})$ | **0.134**  $(4.3*10^{-05})$ |
| $\Delta W_t$ | $-0.043$ $(8.8*10^{-02})$ | $0.026$  $(3.5*10^{-01})$ | **$-$0.309** $(5.4*10^{-32})$ | **0.221**  $(7.5*10^{-14})$ |
| $\Delta S_t$ | **0.316**  $(1.5*10^{-30})$ | $0.054$  $(7.3*10^{-02})$ | $-0.007$  $(7.9*10^{-01})$ | **0.126**  $(4.3*10^{-05})$ |

Vector Auto-Regression results for $P_t$ as price on Mt. Gox, $W_t$ as tweet ratio, $S_t$ as Wikipedia views, and $U_t$ as number of client downloads.

|          | $\Delta P_{t-1}$ | $\Delta U_{t-1}$ | $\Delta W_{t-1}$ | $\Delta S_{t-1}$ |
|----------|------------------|------------------|------------------|------------------|
| $\Delta P_t$ | **0.162**  $(2.0*10^{-08})$ | **0.151**  $(1.0*10^{-05})$ | **0.086**  $(2.9*10^{-03})$ | **$-$0.232** $(3.0*10^{-11})$ |
| $\Delta U_t$ | **0.194**  $(2.7*10^{-11})$ | **$-$0.136** $(7.6*10^{-05})$ | $0.055$  $(5.8*10^{-02})$ | **0.149**  $(1.9*10^{-05})$ |
| $\Delta W_t$ | **$-$0.148** $(1.4*10^{-08})$ | **$-$0.147** $(1.7*10^{-06})$ | **$-$0.446** $(6.0*10^{-59})$ | **0.303**  $(1.8*10^{-21})$ |
| $\Delta S_t$ | **0.373**  $(5.7*10^{-43})$ | $0.006$  $(8.4*10^{-01})$ | **$-$0.079** $(2.6*10^{-03})$ | **0.303**  $(2.6*10^{-21})$ |

Vector Auto-Regression results for $P_t$ as price on Mt. Gox, $W_t$ as Facebook reshares, $S_t$ as Google search volume, and $U_t$ as number of client downloads.

|          | $\Delta P_{t-1}$ | $\Delta U_{t-1}$ | $\Delta W_{t-1}$ | $\Delta S_{t-1}$ |
|----------|------------------|------------------|------------------|------------------|
| $\Delta P_t$ | **0.161**  $(1.4*10^{-08})$ | **0.089**  $(3.7*10^{-03})$ | **0.096**  $(7.1*10^{-04})$ | **$-$0.172** $(7.1*10^{-09})$ |
| $\Delta U_t$ | **0.143**  $(1.1*10^{-07})$ | $-0.112$ $(1.2*10^{-04})$ | $-0.028$  $(2.9*10^{-01})$ | **0.110**  $(8.9*10^{-05})$ |
| $\Delta W_t$ | $-0.033$ $(1.9*10^{-01})$ | $0.007$  $(7.9*10^{-01})$ | **$-$0.320** $(1.0*10^{-33})$ | **0.242**  $(3.8*10^{-19})$ |
| $\Delta S_t$ | **0.384**  $(1.7*10^{-46})$ | $0.040$  $(1.4*10^{-01})$ | $-0.016$  $(5.3*10^{-01})$ | **0.295**  $(5.7*10^{-27})$ |

Vector Auto-Regression results for $P_t$ as price on Mt. Gox, $W_t$ as tweet ratio, $S_t$ as Google search volume, and $U_t$ as new number of users detected in the blockchain.

|          | $\Delta P_{t-1}$ | $\Delta U_{t-1}$ | $\Delta W_{t-1}$ | $\Delta S_{t-1}$ |
|----------|------------------|------------------|------------------|------------------|
| $\Delta P_t$ | **0.167**  $(5.0*10^{-09})$ | $0.040$  $(1.8*10^{-01})$ | **0.099**  $(5.3*10^{-04})$ | **$-$0.163** $(4.1*10^{-08})$ |
| $\Delta U_t$ | **0.129**  $(1.5*10^{-06})$ | **$-$0.235** $(4.3*10^{-16})$ | $-0.030$  $(2.5*10^{-01})$ | **0.108**  $(1.1*10^{-04})$ |
| $\Delta W_t$ | $-0.034$ $(1.7*10^{-01})$ | $0.017$  $(5.1*10^{-01})$ | **$-$0.321** $(7.6*10^{-34})$ | **0.240**  $(4.1*10^{-19})$ |
| $\Delta S_t$ | **0.384**  $(1.4*10^{-46})$ | $0.038$  $(1.6*10^{-01})$ | $-0.016$  $(5.2*10^{-01})$ | **0.297**  $(1.9*10^{-27})$ |

Vector Auto-Regression results for $P_t$ as price on Mt. Gox, $W_t$ as tweet ratio, $S_t$ as Google search volume, and $U_t$ as new number of addresses in the blockchain.

Table S3: Vector Auto-Regression results with alternative metrics of the variables U, W, and S.

# S6    VAR results for other markets and currencies

|          | $\Delta P_{t-1}$                   | $\Delta U_{t-1}$                   | $\Delta W_{t-1}$                   | $\Delta S_{t-1}$                   |
|----------|-----------------------------------|-----------------------------------|-----------------------------------|-----------------------------------|
| $\Delta P_t$ | **0.188**    $(3.9*10^{-08})$ | **0.163**    $(5.1*10^{-05})$ | **0.249**    $(4.8*10^{-12})$ | **−0.356** $(2.4*10^{-16})$ |
| $\Delta U_t$ | **0.222**    $(5.2*10^{-10})$ | −0.057 $(1.7*10^{-01})$ | −0.037   $(3.1*10^{-01})$ | **0.205**    $(4.3*10^{-06})$ |
| $\Delta W_t$ | **−0.103** $(2.0*10^{-03})$ | −0.004 $(9.0*10^{-01})$ | **−0.116** $(8.5*10^{-04})$ | **0.398**    $(1.9*10^{-20})$ |
| $\Delta S_t$ | **0.404**    $(2.2*10^{-35})$ | **0.100**   $(6.0*10^{-03})$ | −0.060  $(6.2*10^{-02})$ | **0.315**    $(1.7*10^{-15})$ |

**Vector Auto-Regression results for $P_t$ as price on Mt. Gox in EUR, $W_t$ as tweet ratio, $S_t$ as Google search volume, and $U_t$ as number of client downloads.**

|          | $\Delta P_{t-1}$                   | $\Delta U_{t-1}$                   | $\Delta W_{t-1}$                   | $\Delta S_{t-1}$                   |
|----------|-----------------------------------|-----------------------------------|-----------------------------------|-----------------------------------|
| $\Delta P_t$ | **0.306**    $(2.5*10^{-21})$ | **0.170**    $(7.2*10^{-06})$ | **0.082**    $(9.6*10^{-03})$ | **−0.352** $(1.8*10^{-19})$ |
| $\Delta U_t$ | **0.249**    $(2.0*10^{-13})$ | **−0.093** $(2.0*10^{-02})$ | −0.018   $(5.7*10^{-01})$ | **0.164**    $(5.4*10^{-05})$ |
| $\Delta W_t$ | **−0.121** $(1.6*10^{-04})$ | 0.052    $(1.7*10^{-01})$ | **−0.275** $(5.3*10^{-17})$ | **0.247**    $(2.7*10^{-10})$ |
| $\Delta S_t$ | **0.478**    $(1.6*10^{-54})$ | 0.046    $(1.7*10^{-01})$ | −0.045  $(1.1*10^{-01})$ | **0.252**    $(6.5*10^{-13})$ |

**Vector Auto-Regression results for $P_t$ as price on BTC China, $W_t$ as tweet ratio, $S_t$ as Google search volume, and $U_t$ as number of client downloads.**

|          | $\Delta P_{t-1}$                   | $\Delta U_{t-1}$                   | $\Delta W_{t-1}$                   | $\Delta S_{t-1}$                   |
|----------|-----------------------------------|-----------------------------------|-----------------------------------|-----------------------------------|
| $\Delta P_t$ | **0.169**    $(1.5*10^{-06})$ | **0.149**    $(3.1*10^{-04})$ | **0.139**    $(1.1*10^{-04})$ | **−0.353** $(1.7*10^{-15})$ |
| $\Delta U_t$ | **0.259**    $(8.8*10^{-13})$ | **−0.087** $(3.8*10^{-02})$ | **−0.090** $(1.4*10^{-02})$ | **0.235**    $(1.6*10^{-07})$ |
| $\Delta W_t$ | **−0.184** $(4.7*10^{-08})$ | 0.027    $(4.7*10^{-01})$ | **−0.083** $(1.5*10^{-02})$ | **0.366**    $(1.0*10^{-17})$ |
| $\Delta S_t$ | **0.439**    $(1.2*10^{-40})$ | 0.056    $(1.2*10^{-01})$ | **−0.151** $(2.9*10^{-06})$ | **0.360**    $(1.3*10^{-19})$ |

**Vector Auto-Regression results for $P_t$ as price on BTC-de, $W_t$ as tweet ratio, $S_t$ as Google search volume, and $U_t$ as number of client downloads.**

Table S4: Vector Auto-Regression results with alternative exchange markets and currencies to measure P.

## S6.1 VAR applied over different periods

| **P1,P2** | $\Delta P_{t-1}$ | $\Delta U_{t-1}$ | $\Delta W_{t-1}$ | $\Delta S_{t-1}$ |
|---|---|---|---|---|
| $\Delta P_t$ | **0.317** $(5.5 * 10^{-21})$ | $-$**0.150** $(2.1 * 10^{-03})$ | $-0.016$ $(6.1 * 10^{-01})$ | **0.146** $(2.3 * 10^{-03})$ |
| $\Delta U_t$ | $-0.005$ $(8.6 * 10^{-01})$ | $-$**0.169** $(5.9 * 10^{-04})$ | $0.005$ $(8.7 * 10^{-01})$ | $-$**0.118** $(1.5 * 10^{-02})$ |
| $\Delta W_t$ | $-0.011$ $(7.0 * 10^{-01})$ | $0.006$ $(8.8 * 10^{-01})$ | $-$**0.372** $(6.1 * 10^{-34})$ | **0.085** $(4.9 * 10^{-02})$ |
| $\Delta S_t$ | $-$**0.099** $(2.5 * 10^{-03})$ | **0.122** $(1.2 * 10^{-02})$ | $0.015$ $(6.2 * 10^{-01})$ | $-$**0.377** $(1.2 * 10^{-14})$ |

| **P3** | $\Delta P_{t-1}$ | $\Delta U_{t-1}$ | $\Delta W_{t-1}$ | $\Delta S_{t-1}$ |
|---|---|---|---|---|
| $\Delta P_t$ | **0.176** $(1.0 * 10^{-03})$ | **0.163** $(9.7 * 10^{-03})$ | **0.283** $(1.0 * 10^{-06})$ | $-$**0.329** $(1.6 * 10^{-06})$ |
| $\Delta U_t$ | **0.220** $(7.9 * 10^{-05})$ | $-0.030$ $(6.4 * 10^{-01})$ | $-0.023$ $(6.8 * 10^{-01})$ | **0.188** $(7.1 * 10^{-03})$ |
| $\Delta W_t$ | $-0.031$ $(5.4 * 10^{-01})$ | $-0.007$ $(9.0 * 10^{-01})$ | $-0.063$ $(2.4 * 10^{-01})$ | **0.431** $(1.0 * 10^{-10})$ |
| $\Delta S_t$ | **0.382** $(2.9 * 10^{-14})$ | **0.117** $(3.8 * 10^{-02})$ | $-0.038$ $(4.5 * 10^{-01})$ | **0.302** $(9.4 * 10^{-07})$ |

Table S5: Vector autoregression results for $P_t$ as price on Mt. Gox, $S_t$ as Google search volume, and $U_t$ as identified users in the block chain, for the period between the opening of Mt. Gox and the end of the second bubble (P1+P2), and the third bubble (P3).

## S6.2 Extended VAR results

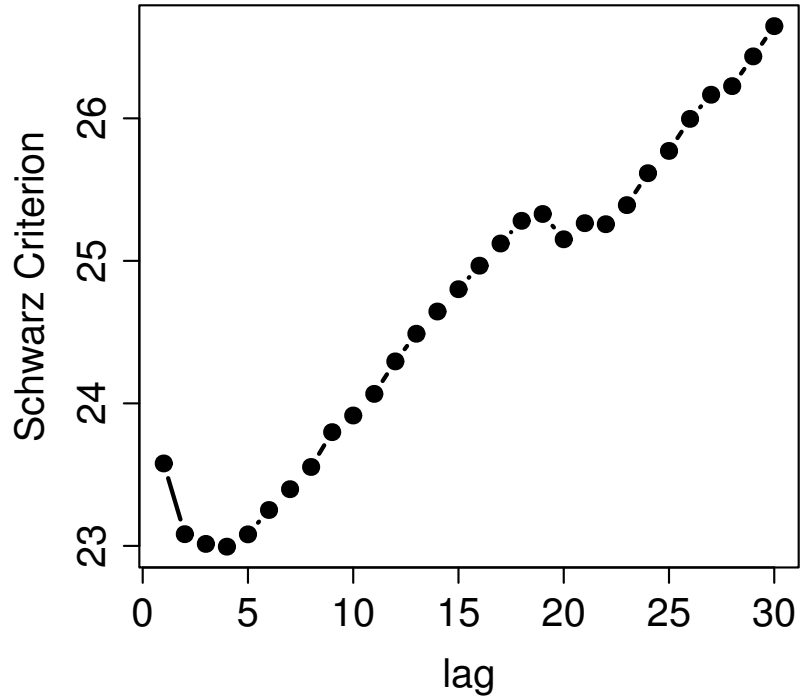

Figure S4: Schwarz Bayesian information criterion over VAR of lags from 1 to 30. The Schwarz criterion reaches its minumum at lag = 4 days.

| Price increase | 33.18 | 27.23 | 24.64 | 22.05 | 20.28 | 19.94 | 17.71 | 17.44 | 15.37 | 14.13 |
|---|---|---|---|---|---|---|---|---|---|---|
| VAR estimate | 14.88 | 12.41 | 12.26 | 17.56 | 1.18 | 23.95 | 4.13 | 10.17 | 1.46 | 6.30 |
| Price decrease | -73.36 | -30.02 | -25.72 | -21.74 | -18.82 | -16.90 | -15.19 | -14.56 | -14.50 | -12.80 |
| VAR estimate | -48 | 3.54 | -18.57 | -21.05 | -1.47 | -0.49 | -3.96 | -7.38 | -2.23 | -6.41 |

Table S6: Top 10 strongest price increases and top 10 strongest price decreases between November 30th, 2012 and October 30th, 2013, and VAR estimate with lag 4. The VAR correctly estimates the sign of the top 10 increases and of 9 out of the top 10 decreases.

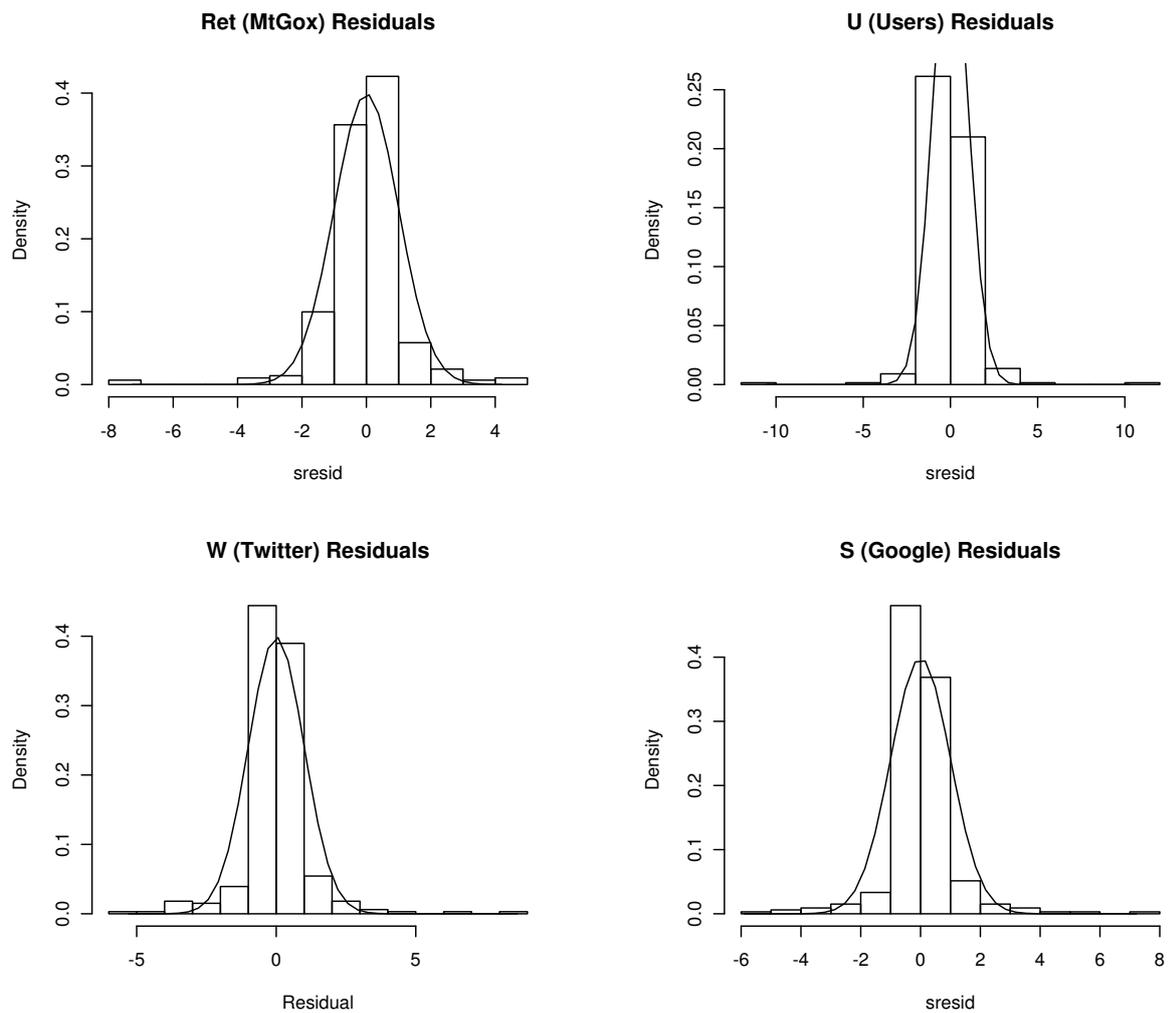

Figure S5: Distributions for the Studentised residuals of the four estimates of the model, which resemble normal distributions. Nevertheless, Shapiro-Wilk tests reject the normal distribution, indicating that improvements of the model are possible, especially for the outliers.

# S7 Detailed results of impulse response functions

| shock | $P_{t+1}$ | $P_{t+2}$ | $S_{t+1}$ | $S_{t+2}$ | $W_{t+1}$ | $W_{t+2}$ | $U_{t+1}$ | $U_{t+2}$ |
|-------|-----------|-----------|-----------|-----------|-----------|-----------|-----------|-----------|
| $P_t$ | 0.94 | 0.14 | – | -0.139 | – | 0.087 | 0.175 | 0.142 |
| $S_t$ | – | 0.374 | 0.845 | 0.281 | – | – | – | – |
| $W_t$ | – | – | – | 0.192 | 0.86 | -0.277 | – | – |
| $U_t$ | – | 0.175 | 0.517 | – | – | – | 0.818 | -0.086 |

Table S7: Response function estimates of lags 1 and 2 under 97.5% confidence level (95% after Bonferroni correction).



\end{document}